\documentclass[journal]{IEEEtran}

\ifCLASSINFOpdf
\else
   \usepackage[dvips]{graphicx}
\fi
\usepackage{url}
\usepackage{amsmath}  
\usepackage{amssymb} 
\usepackage{multirow}
\usepackage{makecell}  
\usepackage{subcaption} 
\usepackage{float}  
\usepackage{balance} 
\usepackage{cite}  

\usepackage{graphicx}

\begin{document}

\title{\huge Time-Variance Aware Real-Time Speech Enhancement}

\author{Chengyu\ Zheng$^{1*}$, Yuan\ Zhou$^{2}$, Xiulian\ Peng$^{2}$, Yuan\ Zhang$^{1}$ and Yan\ Lu$^{2}$ \\ 
$^{1}$Communication University of China, Beijing, China \\
$^{2}$Microsoft Research Asia, Beijing, China \\
\thanks{*This work was done when Chengyu Zheng was an intern at Microsoft Research Asia.}}

\maketitle

\begin{abstract}
Time-variant factors often occur in real-world full-duplex communication applications. Some of them are caused by the complex environment such as non-stationary environmental noises and varying acoustic path while some are caused by the communication system such as the dynamic delay between the far-end and near-end signals. 
Current end-to-end deep neural network (DNN) based methods usually model the time-variant components implicitly and can hardly handle the unpredictable time-variance in real-time speech enhancement. To explicitly capture the time-variant components, we propose a dynamic kernel generation (DKG) module that can be introduced as a learnable plug-in to a DNN-based end-to-end pipeline. Specifically, the DKG module generates a convolutional kernel regarding to each input audio frame, so that the DNN model is able to dynamically adjust its weights according to the input signal during inference. Experimental results verify that DKG module improves the performance of the model under time-variant scenarios, in the joint acoustic echo cancellation (AEC) and deep noise suppression (DNS) tasks. 
\end{abstract}

\begin{IEEEkeywords}
speech enhancement, acoustic echo cancellation, deep noise suppression, time-variance, deep neural network
\end{IEEEkeywords}

\IEEEpeerreviewmaketitle

\section{Introduction}

\IEEEPARstart{A}{udio} signals are usually interfered during the real-time communications, which lead to the degradation of the speech quality and the user experiences. On the sender side of the audio communication pipeline, environmental noise and acoustic echo are the main interfering factors to affect the quality of the near-end speech. Echo occurs due to coupling of the loudspeaker and the microphone in a real-time communication system such that the user at the far end hears a delayed and modified version of his/her own voice. Therefore, speech enhancement including deep noise suppression (DNS) and acoustic echo cancellation (AEC), aims at removing both the echo and environmental noises and transmitting only the near-end speech to the far-end.

Time-variant factors often occur in real-time full-duplex communication applications. User movement or environmental changes may lead to the varying acoustic path and non-stationary noises. The frontend signal transmission and pre-processing modules often bring the frame-wise misalignment between the microphone and far-end signals. This may lead to the time delay between the dual signals changing along with the audio frames, which we term as ``dynamic delay". 
In conventional speech enhancement algorithms, these time-variant components are captured via dynamically tracking the input signal in an adaptive way \cite{kao2003design,nathwani2018joint,djendi2019new}. 

Recent works regard speech enhancement as a time-series regression problem and use deep neural network (DNN) for its powerful capacity of nonlinear modeling, which can be divided into DSP-DNN hybrid methods \cite{shu2021joint,valin2021low,peng2021acoustic,gu2021residual,franzen2022deep,sun2022explore,zhang2022multi,wang2022nn3a} and end-to-end DNN methods \cite{watcharasupat2022end,yu2022neuralecho,zheng21ft_interspeech,cui2022multi}. 
In the hybrid methods, the DSP module explicitly captures the time-variance and partially suppresses the echo and noise, while the DNN module works as a post-processor to cancel the residual interferences. In the end-to-end methods, the DNN can also model the time-variant components but in an implicit way. Inspired by the both kinds of methods, empowering the DNN with the explicit time-variance awareness and modeling capacity may reinforce its performance on processing time-variant signals, especially in an end-to-end way.

In this paper, we propose a dynamic kernel generation (DKG) module for explicitly modeling the time-variance in real-time speech enhancement. This DKG module can be introduced as a learnable plug-in and trained with the end-to-end optimization of DNN. Specifically, with each input audio frame, the DKG module generates a convolutional kernel and applies it to the features of both the current and historical audio frames, then the recalibrated features are used to get the corresponding output audio frame. This enables the DNN model to dynamically adjust its weights according to the time-variant inputs during the inference. 
We introduce two different structures of DKG, i.e., separable and non-separable DKG, for different implementations of the time-variant components capturing. Ablation studies on the synthetic dataset show that the proposed DKG module improves the model performance especially under the time-variant scenarios including varying acoustic path and dynamic delay. Experimental results on the real-world dataset also verify the effectiveness of the proposed module. 

\section{Proposed Methods}
\label{section_2}

\subsection{Problem Formulation}

In the conventional acoustic signal model, the microphone signal $y(t)$ is the mixture of near-end signal $s(t)$, echo $d(t)$ and the background noise $n(t)$:
\begin{equation}
    y(t)=d(t)+s(t)+n(t).
\end{equation}

The echo signal $d(t)$ is generated from the far-end signal by first distorted by the nonlinear components e.g., the power amplifier and the loudspeaker, and then convolved with a room impulse response (RIR). The joint AEC and DNS problem is to estimate the clean near-end signal from the microphone signal. 

\subsection{Model Architecture}

\begin{figure}
\centerline{\includegraphics[width=\columnwidth]{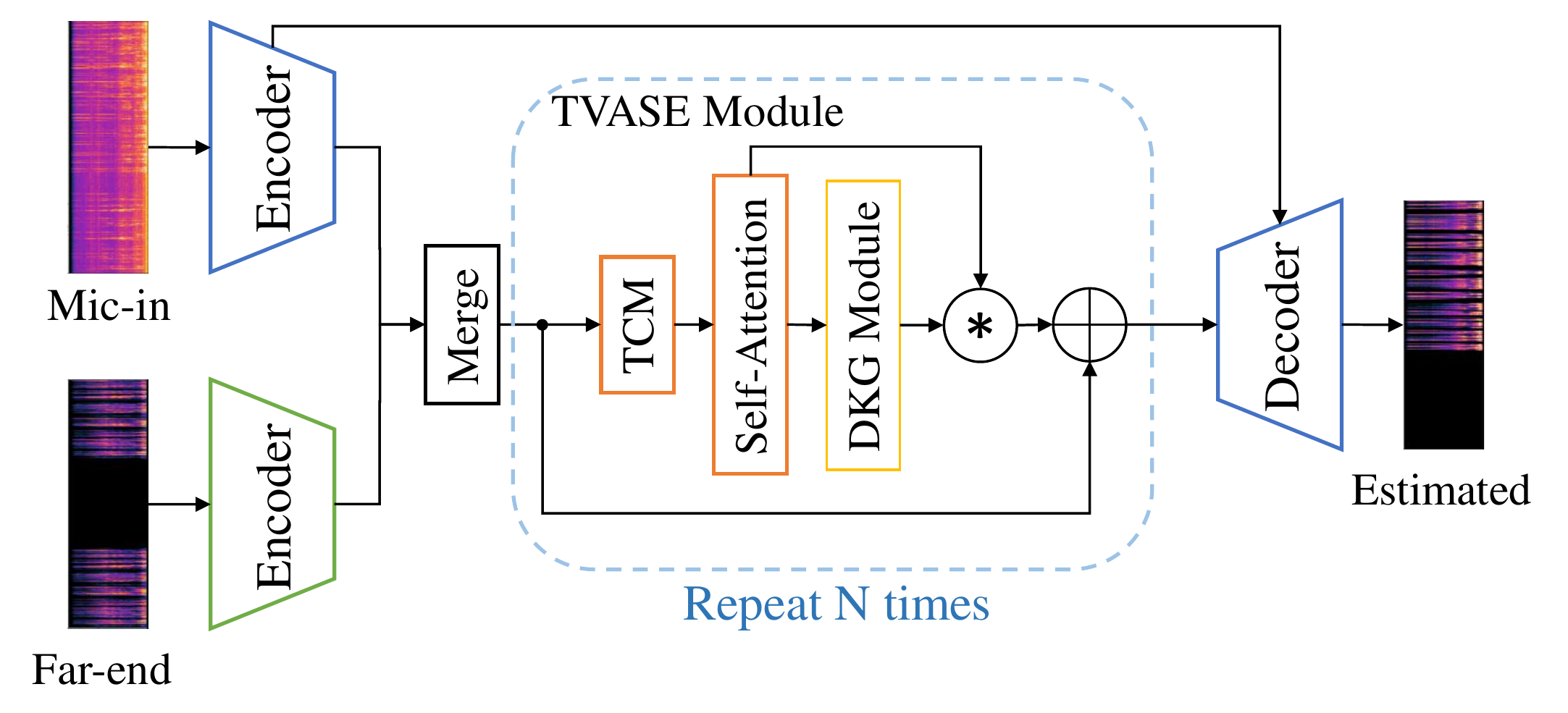}}
\caption{Overall architecture of the joint model with the proposed DKG module.}
\label{fig_overall}
\end{figure}

Fig. \ref{fig_overall} shows the overall architecture of the joint model with the proposed DKG module. The model consists of two individual encoders, one decoder and several repeated time-variance aware speech enhancement (TVASE) modules connecting between them. The model takes the Short-time Fourier Transform (STFT) spectrum of the microphone $\mathcal{Y}\in \mathbb{R}^{2\times T\times F}$ and the far-end $\mathcal{X}\in \mathbb{R}^{2\times T\times F}$ signals as the inputs and estimates the STFT spectrum of the near-end signal, where $T$ is the frame number, $F$ is the number of frequency bins and each complex spectrum has real and imaginary parts.

The microphone and the far-end spectrum are input to two individual encoders, respectively. Each encoder contains four 2-D causal convolutional layers \cite{krizhevsky2012imagenet}, which gradually down-sample the feature along the frequency dimension and increase the number of its channels. The features output from the two encoders are concatenated and fed into a 2-D causal convolutional layer, and then the frequency dimension is merged to the channel dimension to get the final encoder feature of shape $\mathbb{R}^{(F'C)\times T}$. 

The TVASE module contains a temporal convolution module (TCM) defined in \cite{pandey2019tcnn}, a self-attention module and a DKG module, as depicted in the dotted box in Fig. \ref{fig_overall}. The incorporation of the TCM and the self-attention module aims at capturing local and global dependencies along the temporal dimension simultaneously, while the DKG module focuses on modeling the time-variant components of the input features explicitly. 

Inspired by the design of multi-head self-attention which extracts information from different subspace of the features \cite{vaswani2017attention}, we split the features from the TCM $\mathcal{F}_{R}\in\mathbb{R}^{(F'C)\times T}$ into $I$ groups and get $\{\mathcal{F}_{i}=\mathcal{F}_{R}[iC':(i+1)C',:],i\in \{0,...,I-1\},C'=\frac{F'C}{I}\}$. The scaled dot-product self-attention is conducted on each group. All groups of features $\mathcal{F}_i^{SA}$ are concatenated along the channel dimension and fed into a convolutional layer to obtain the feature $\mathcal{F}^{SA}\in\mathbb{R}^{(F'C)\times T}$. A windowed mask with window size of $T_{w}$ is applied inside the self-attention to keep its causality:
\begin{equation}
    \begin{aligned}
        &\mathcal{F}_{i}^{k} = (Conv_{i}^{k}(\mathcal{F}_{i}))^{tr},k\in\{K,Q,V\}, \\
        &\mathcal{F}_{i}^{SA} = (Softmax(Mask(\mathcal{F}_{i}^{Q}\cdot(\mathcal{F}_{i}^{K})^{tr}/\sqrt{C'}))\cdot \mathcal{F}_{i}^{V})^{tr}, \\
        &Mask(x)(i,j) = \left\{
        \begin{array}{rcl}
        x(i,j), & & 0\le i-j\le T-T_w \\
        -\infty, & & otherwise \\
        \end{array}
        \right.
    \end{aligned}
\end{equation}
where $\mathcal{F}_{i}\in\mathbb{R}^{C'\times T}$, $\mathcal{F}_{i}^{k}\in\mathbb{R}^{T\times C'}$, and $\mathcal{F}_{i}^{SA}\in\mathbb{R}^{C'\times T}$, respectively. The superscription $tr$ means transpose the last two dimensions of the tensor. $Conv^{k}_{i}$ represents a 1-D convolutional layer followed by a batch normalization (BN) \cite{ioffe2015batch} and a parametric ReLU (PReLU) \cite{he2015delving}. 

The decoder consists of four gated blocks similar to \cite{zheng2021interactive} but with causal convolutions and an extra 2-D causal convolutional layer at last. 
Except the last layer in the decoder, all the other convolutional layers are followed by BN and PReLU.

\subsection{DKG Module}

To better capture time-variant components including varying acoustic path and dynamic delay, we introduce the DKG module to enable the model to adapt its weights according to the input signal in the inference phase.

Given the input feature $\mathcal{F}^{SA}\in\mathbb{R}^{(F'C)\times T}$ and the kernel size $M$, DKG module generates a convolutional kernel $\mathcal{K}\in\mathbb{R}^{(F'C)\times T\times M}$ regarding the input feature. Then for each channel of a single feature frame $\mathcal{F}^{SA}(c,t)$, the kernel $\mathcal{K}(c,t)\in\mathbb{R}^{M}$ is applied to get the corresponding output:

\begin{equation}
    \begin{aligned}
        \mathcal{F}^{O}(c,t)=\sum_{t'=t-t_0}^{t}\mathcal{K}(c,t,t'-(t-t_0))\mathcal{F}^{SA}(c,t'),\\
    \end{aligned}   
\end{equation}
where $t_0=M-1$, $c\in\{0,...,F'C-1\}$ and $t\in\{0,...,T-1\}$.

\begin{figure}
\centering
\begin{subfigure}[b]{0.8\columnwidth}
    \centerline{\includegraphics[width=\textwidth]{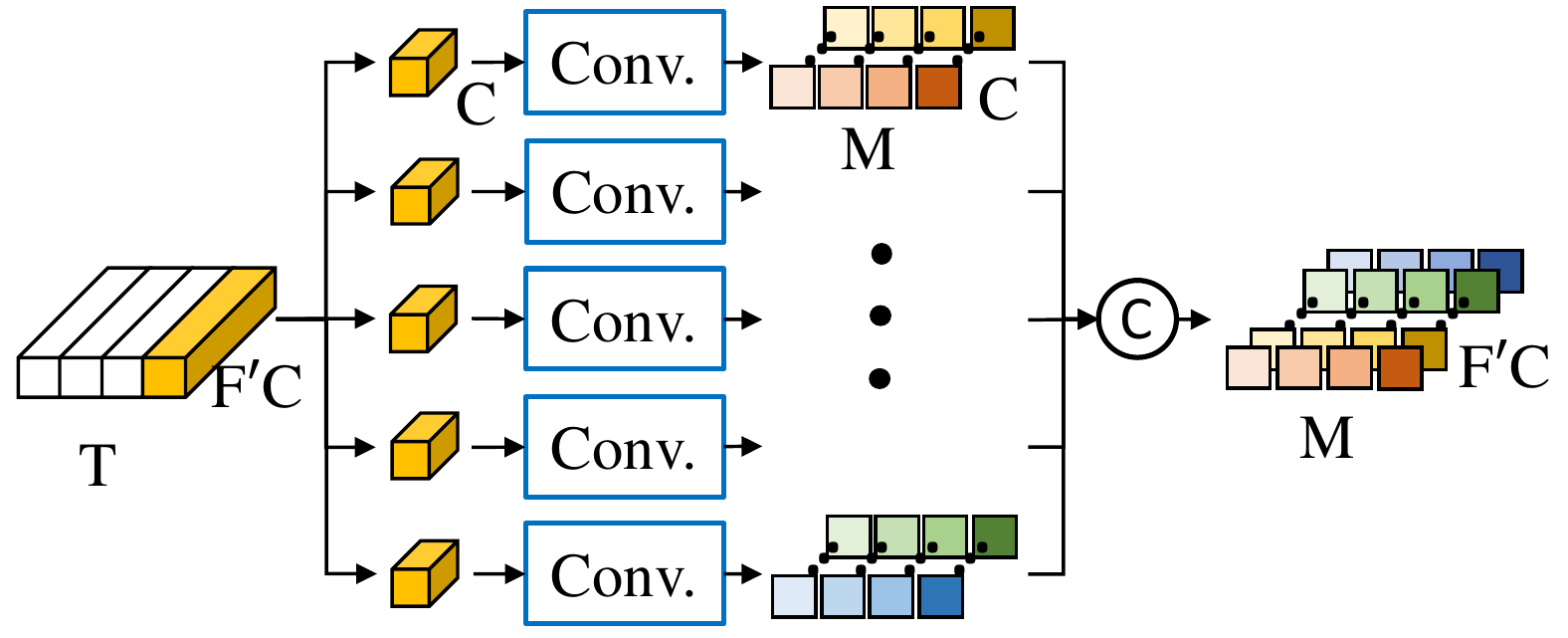}}
    \caption{}
    \label{fig_kg_non_separable}
\end{subfigure}
\hfill
\begin{subfigure}[b]{0.8\columnwidth}
    \centerline{\includegraphics[width=\textwidth]{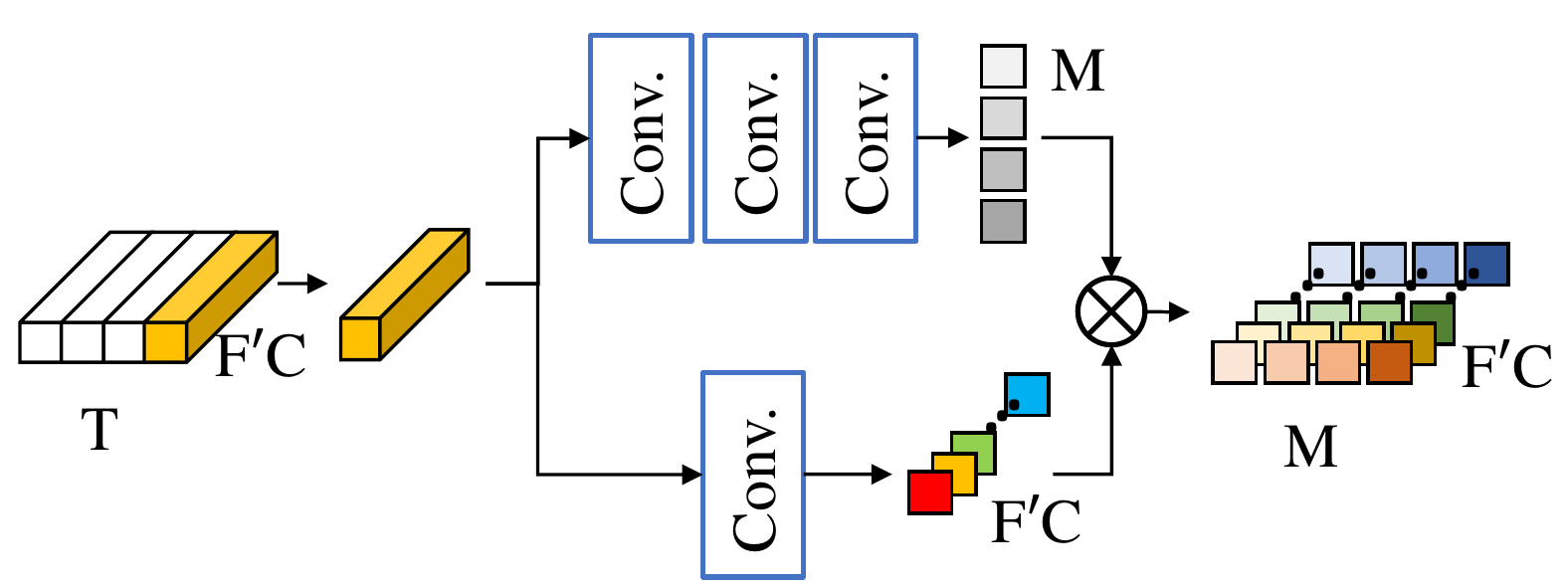}}
    \caption{}
    \label{fig_kg_separable}
\end{subfigure}
\caption{The structure of dynamic kernel generation module. (a) Non-separable. (b) Separable.}
\end{figure}

Based on whether to generate the informative weights along the temporal and channel dimension separately, we propose two types of structures for DKG module, i.e. non-separable and separable DKG.

The non-separable DKG module is shown in Fig. \ref{fig_kg_non_separable}. In this structure, the kernel is generated using a single mapping directly: $\mathbb{R}^{(F'C)\times T}\mapsto\mathbb{R}^{(F'C)\times T\times M}$. To reduce the complexity,
we split the input feature $\mathcal{F}^{SA}$ into $F'$ groups $\{\mathcal{F}^{SA}_{i}=\mathcal{F}^{SA}[iC:(i+1)C,:]\},i\in\{0,...,F'-1\},\mathcal{F}^{SA}_{i}\in\mathbb{R}^{C\times T}$. For each feature group, a 1-D convolutional layer is used to generate the kernel $\mathcal{K}^{SA}_i(m)$: $\mathbb{R}^{C}\mapsto\mathbb{R}^{C\times1\times M}$. Then, all groups of kernels are concatenated along the channel dimension to get the kernel $\mathcal{K}$.

The separable DKG is shown Fig. \ref{fig_kg_separable}.  
In this structure, the kernel is generated using two separated mappings, including one to generate a channel-sharing filter $\mathcal{K}_0$ and the other to generate a channel-dependent weight $\mathcal{K}_s$ that is then multiplied to $\mathcal{K}_0$ element-wise. For each audio frame, the channel-sharing filter is generated using three 1-D convolutional layers: $\mathbb{R}^{F'C}\mapsto\mathbb{R}^{1\times1\times M}$, and the channel-dependent weight is generated using a 1-D convolutional layer: $\mathbb{R}^{F'C}\mapsto\mathbb{R}^{(F'C)\times1\times1}$. 

\section{Experiment Settings}

\subsection{Training Datasets}

\begin{table}
\centering
\caption{Detailed description of the synthetic test set}
\small
\setlength{\tabcolsep}{3pt}
\resizebox{\columnwidth}{!}{
    \begin{tabular}{l|c|c|c}
        \hline
        Scenarios& Varying RIR& Dynamic Delay & Delay Range (ms) \\ 
        \hline
        Time-invariant& $\times$& $\times$& $[0,100]$ \\
        \hline
        Variant-delay-only& $\times$ & $\checkmark$ & $[0,100]\pm[-20,20]$ \\
        \hline
        Variant-RIR-only& $\checkmark$ & $\times$ & $[0,100]$ \\
        \hline
        Variant-delay-and-RIR& $\checkmark$ & $\checkmark$ & $[0,100]\pm[-20,20]$ \\
        \hline
    \end{tabular}
    }
\label{table_synt_set}
\end{table}

We synthesize 500 hours of audio samples for training and 8 hours for validation. The far-end, near-end and the noise signals are all from DNS challenge data at Interspeech 2021 \cite{reddy21_interspeech}. The RIRs are from the AEC challenge data at Interspeech 2021 \cite{cutler21_interspeech}. We convolve the far-end signal with a randomly chosen RIR to generate the echo signal. In 80$\%$ of the cases, the far-end signal is nonlinearly distorted at first, by subsequently performing the hard clipping to simulate the characteristic of a power amplitude and applying the sigmoidal function to simulate the loudspeaker distortion \cite{zhang19o_interspeech}.

A time delay uniformly sampled from 0 to 900 ms is applied to get the echo signals. Finally, the microphone signal is generated by mixing the near-end signal with the noise and the echo signal at an SNR uniformly sampled from -5 dB to 20 dB and an SER uniformly sampled from -15 dB to 15 dB, respectively.

We use both AEC and DNS test sets for validating the effectiveness of the models on the joint speech enhancement tasks. We will introduce the details of each task respectively.
\subsection{AEC Test Sets}
Two test sets including a synthetic test set for ablation study and a real-recorded test set for fair and reproducible comparison between the models. For the synthetic test set, we use TIMIT dataset \cite{lamel1989speech} as the source data and follow the steps reported in \cite{zhang19o_interspeech} to synthesize 300 far-end and near-end signal pairs.

50$\%$ of the far-end signals are nonlinearly distorted. The RIRs are generated using the image method \cite{allen1979image}. We simulate 60 different rooms in the size of $a\times b\times c$, where $a\in\{5,7,9,11,13\}$, $b\in\{4,6,8,10\}$ and $c\in\{2.5,3.5,4.5\}$, with the loudspeaker fixed at the center of the room. The $RT_{60}$ ranges from 0.3s to 1.3s. A basic delay in the range of 0 to 100 ms is added to each far-end signal.

We manually introduce the time-variant factors, i.e. the varying acoustic path and dynamic delay, to the synthetic test set. To mimic the varying acoustic path, we first generate a group of 400 continuously varying RIRs by changing the relative positions between the microphone and loudspeaker. In each room, one microphone starts from the position of the loudspeaker and keeps moving to 400 different positions continuously with the moving step $(\Delta a,\Delta b)$, where $\Delta a,\Delta b\in[-0.025,0.025]$. The symbols of $\Delta a$ and $\Delta b$ will not change until the microphone reaches the border of the room. Then a series of continuous RIRs are randomly selected from the 400 RIRs and applied to the far-end signal at intervals of 500 ms. To mimic the dynamic delay, an extra varying delay ranging from -20 ms to 20 ms is added to the far-end signal every 500 ms, in addition to the basic delay.

Finally, we synthesize 4 test scenarios including time-invariant, variant-delay-only, variant-RIR-only, and variant-delay-and-RIR with the different RIRs and delays shown in Table \ref{table_synt_set}. 

Each scenario contains 900 pairs of microphone and far-end signals, resulting from 300 pairs of far-end and near-end signals with SER of {0, 3.5, 7} dB.

The real-recorded test set is the blind test set of AEC Challenge Interspeech 2021 \cite{cutler21_interspeech}. This test set consists of 800 real world recordings including three talking scenarios: doubletalk, farend-singletalk and nearend-singletalk.
\subsection{DNS Test Sets}
We use the blind test set of Track 1 at DNS Challenge Interspeech 2021 \cite{reddy21_interspeech}. The test set includes utterances recording in the presence of a variety of background noises at different SNR, target levels, acoustic conditions, also covers people talking in different languages, emotions and with musical instruments in the background, to enrich the diversity of the data. All the clips are originally collected at a sampling rate of 48 kHz and resampled to 16 kHz.

\subsection{Implementation Details}

All the training utterances are clipped to 3 seconds. All signals are resampled in 16 kHz and transformed to STFT domain using a 20-ms Hanning window, 10-ms overlap and 320-point Discrete Fourier Transform (DFT). 

The details of the joint model are as follows. For all the convolutional layers in the encoder, the kernel size is (2,5), strides are (1,1), (1,4), (1,4), (1,2) and the output channels are 16, 32, 64, 64, resulting in the output feature of shapes $\mathbb{R}^{16\times T\times 161}$, $\mathbb{R}^{32\times T\times 41}$, $\mathbb{R}^{64\times T\times 11}$, and $\mathbb{R}^{64\times T\times 5}$, respectively. For all the deconvolutional layers in the decoder, the kernel size is (2,5), strides are (1,2), (1,4), (1,4), (1,1) and channels are 64, 32, 16, 2, respectively. For the last convolutional layer in the decoder, we use the kernel size of (2,5), stride of (1,1) and channel of 2. For each convolutional layers of the gated blocks in the decoder, the kernel size is (1,1), stride is (1,1) and the number of channels are the same as the corresponding encoder features. Four TVASE modules are used between the encoders and the decoder. For the TCM in the TVASE module, the kernel size is 1 for the 1-D convolutional layer and 3 for the depth-wise 1-D convolution. All the strides are 1, and the output channels are 256, 256, 320, respectively. The group number $I$ of temporal self-attention is 5. All the convolutional layers of self-attention module have the kernel size of (1,1), stride of (1,1) and channels of 64. The window size $T_{w}$ is 100. For the DKG module, the kernel size and stride of all the convolutional layers are  1. The size of the generated kernel $M$ is 10. For the non-separable DKG, the channel number of all the convolutional layers are 640. For the separable DKG, the channel number of 3 convolutional layers to generate $\mathcal{K}_0$ are 80, 20, 10, respectively, and the one to generate $\mathcal{K}_s$ is 320. 

The mean-square-error loss on the power-law compressed STFT spectrum \cite{ephrat2018looking} is minimized in the training. An inverse STFT and forward STFT are conducted on the output of the model before calculating the loss to ensure STFT consistency \cite{wisdom2019differentiable}. Adam optimizer with a learning rate of 0.0003 is used. All the layers are initialized with Xavier initialization. The proposed algorithm is implemented in PyTorch. The model is trained for 200 epochs with a batch size of 200. The model with the minimum validation loss is selected to evaluate on the test sets.

\begin{table}
\centering
\caption{Number of parameters and complexity of different structures.}
\small
\begin{tabular}{c|c|c}
    \hline
    Models& Para.(M)& MACs/sec.(M)\\ 
    \hline
    Backbone& 1.97& 462.10 \\
    \hline
    Backbone*& 2.64& 527.95 \\
    \hline
    +Non-separable DKG& 2.82& 545.30 \\
    \hline
    +Separable DKG& 2.50& 515.30 \\
    \hline
\end{tabular}
\label{table_param_comp}
\end{table}

\begin{table}
\centering
\caption{Ablation study on AEC.}
\small
\begin{tabular}{c|c|c}
    \hline
    Models& ERLE& PESQ \\ 
    \hline
    Unprocessed& -& $1.483$ \\
    \hline
    Backbone& $27.755\pm 1.222$& $2.305\pm 0.009$ \\
    \hline
    Backbone*& $29.998\pm 0.708$& $2.346\pm 0.026$ \\
    \hline
    +Non-separable DKG& $30.679\pm 2.165$& $2.382\pm 0.024$ \\
    \hline
    +Separable DKG& $\textbf{30.708}\pm\textbf{0.748}$& $\textbf{2.395}\pm\textbf{0.018}$ \\
    \hline
\end{tabular}
\label{table_ablation_aec}
\end{table}

\subsection{Evaluation Metrics}

For the synthetic test set, we use the objective evaluation metrics including echo return loss enhancement (ERLE, only for single-talk periods in AEC), perceptual evaluation of speech quality (PESQ) \cite{rec2005p} 
the AEC performance is evaluated in terms of echo return loss enhancement (ERLE) for the single-talk periods and perceptual evaluation of speech quality (PESQ) \cite{rec2005p} for the double-talk periods. The ERLE is defined as:
\begin{equation}
    ERLE=10log_{10}(\mathbb{E}(y^2(m))/\mathbb{E}(s^2(m)))
\end{equation}

For the real-recorded test sets, we use the AECMOS tool \cite{purin2021aecmos} with regards to echo ratings and other degradation ratings and DNSMOS tool \cite{kareddy2020dnsmos} with regards to noise suppression and speech degradation ratings to evaluate all the methods. 

\section{Experimental Results}

\subsection{Ablation Study}

\begin{figure}
\centerline{\includegraphics[width=0.8\columnwidth]{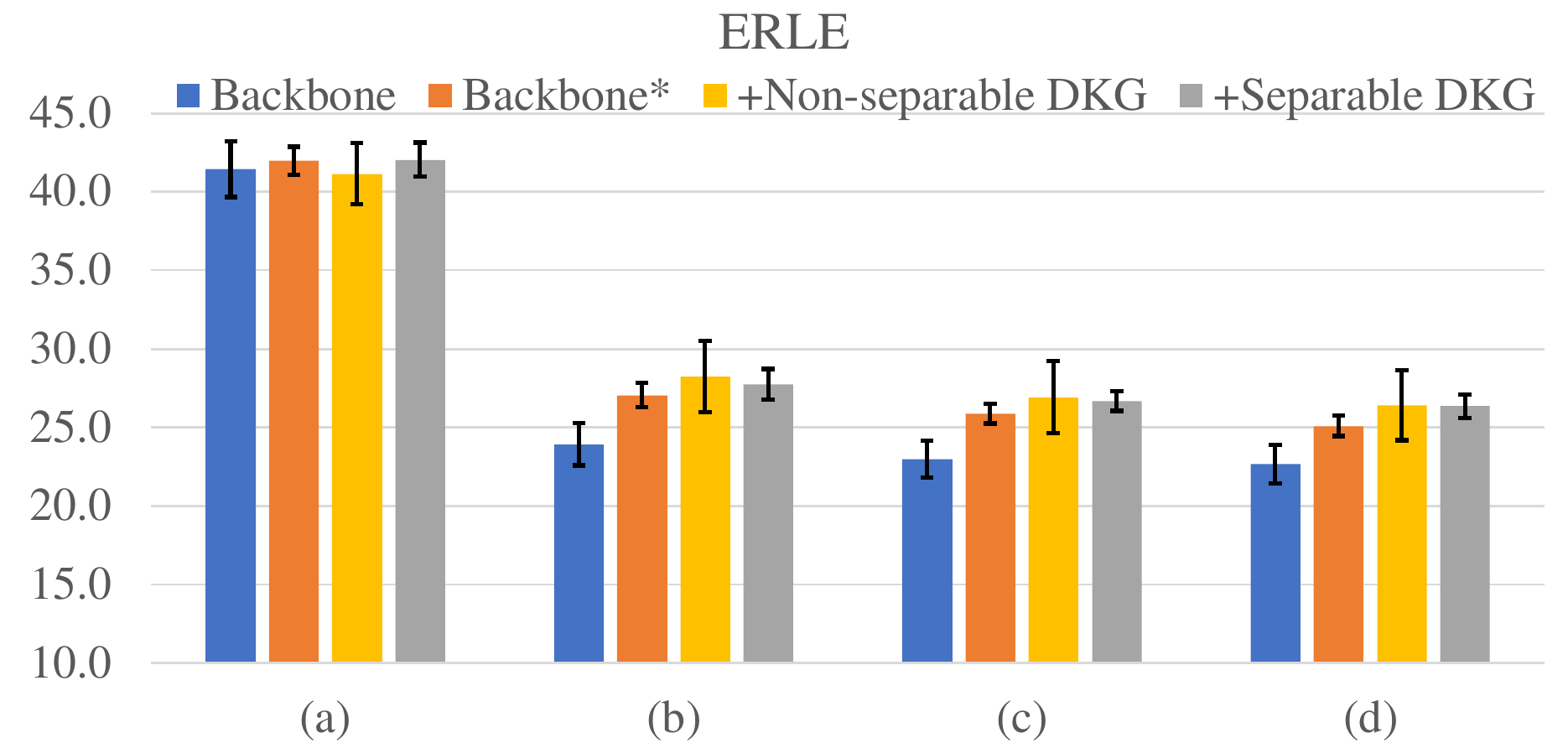}}
\caption{ERLE results of different scenarios on the synthetic test set. (a) Time invariant. (b) variant-delay-only. (c) variant-RIR-only. (d) variant-delay-and-RIR.}
\label{fig_ablation_ERLE}
\end{figure}

\begin{figure}[t]
\centerline{\includegraphics[width=0.8\columnwidth]{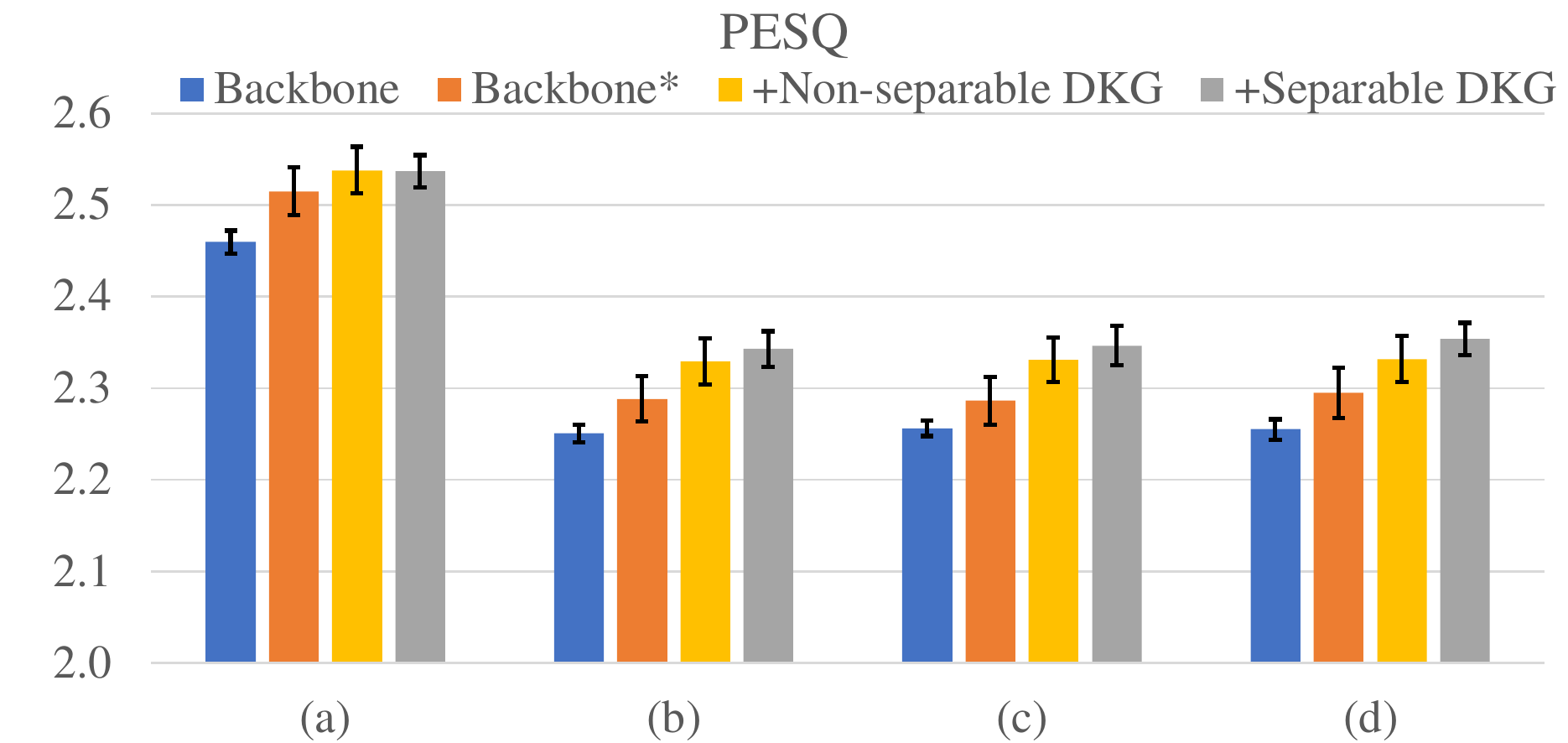}}
\caption{ PESQ results of different scenarios on the synthetic test set. (a) Time invariant. (b) variant-delay-only. (c) variant-RIR-only. (d) variant-delay-and-RIR. The PESQ numbers of the unprocessed signals are 1.548, 1.455, 1.462, 1.464, correspondingly. }
\label{fig_ablation_PESQ}
\end{figure}

Ablation experiments are conducted on four structures. We remove the DKG module as the backbone model and enlarge the model size of the backbone for fair comparison, denoted as Backbone and Backbone*, respectively. Table \ref{table_param_comp} shows the number of parameters and complexity of different structures. 

For AEC, we use the synthetic test set to validate the effectiveness of the proposed module. The results are shown in Table \ref{table_ablation_aec}. We find that introducing the DKG to the TVASE module improves both the ERLE and PESQ. Moreover, the model with separable DKG module slightly outperforms the one with non-separable DKG module. The cross-channel dependency of the separable DKG could bring the global information to the model to better distinguish the signals with similar characteristics, while the non-separable DKG can only capture local patterns, which might lead to the degradations of the non-separable structure on the AEC task, especially for the signals interfered with speech-related characteristics.

Fig. \ref{fig_ablation_ERLE} and Fig. \ref{fig_ablation_PESQ} show the evaluation metrics of different model setups under different time-invariant/variant scenarios. 
For the time-invariant scenario shown in Fig. \ref{fig_ablation_PESQ} (a), the introducing of DKG module improves the PESQ value which indicates the performance on double-talk periods. This shows that DKG module enables the backbone model to better distinguish the targeted speech-related characteristics from the mixing signals. For the scenarios that contain single time-variant factor, i.e., variant-delay-only in Fig. \ref{fig_ablation_ERLE} (b) and Fig. \ref{fig_ablation_PESQ} (b), and variant-RIR-only in Fig. \ref{fig_ablation_ERLE} (c) and Fig. \ref{fig_ablation_PESQ} (c), both the ERLE and PESQ values get improved when the DKG module is introduced to the backbone model. This verifies that DKG module can better capture the time-variant patterns including the dynamic time misalignment and varying acoustic path. Also, as shown in Fig. \ref{fig_ablation_ERLE} (d) and Fig. \ref{fig_ablation_PESQ} (d), when these two time-variant components occur inside one case, the introducing of DKG still improves both metrics, which indicates the robustness of DKG module to handle the complex interwined time-variant pattern. 

\begin{table}
\centering
\caption{Ablation study on DNS.}
\resizebox{\columnwidth}{!}{
    \begin{tabular}{c|c|c|c}
        \hline
        Models& SIG& BAK& OVL \\ 
        \hline
        Unprocessed& \textbf{3.830} & 3.090 & 3.100 \\
        \hline
        Backbone& $3.644\pm 0.009$& $4.282\pm 0.018$& $3.434\pm 0.011$ \\
        \hline
        Backbone*& $3.660\pm 0.007$& $4.302\pm 0.015$& $3.452\pm 0.008$ \\
        \hline
        +Non-separable DKG& $3.712\pm 0.011$& $\textbf{4.322}\pm \textbf{0.013}$& $\textbf{3.508}\pm \textbf{0.013}$ \\
        \hline
        +Separable DKG& $3.692\pm 0.008$& $4.318\pm 0.018$& $3.496\pm 0.011$ \\
        \hline
    \end{tabular}
}
\label{table_ablation_dns}
\end{table}

We also conduct the ablation studies on the DNS test set. The results are shown in Table \ref{table_ablation_dns}. 
Significant improvements are shown on the DNSMOS, including SIG (for signal), BAK (for background) and OVL (for overall). This means the DKG module brings advantages on both suppressing the background noise and keeping the fidelity of the foreground speech, which is usually a pair of mutually exclusive tasks in the speech enhancement. 
The obsevation is similar to the AEC task. Non-separable DKG outperforms separable DKG, which indicates that local patterns of speech and noise are effective and important for DNS.

\subsection{Comparison with Other Methods}

\begin{table}
\centering
\caption{Comparison with other methods on the blind test set at Interspeech 2021. FST, DT, and NST represent the far-end single-talk, double-talk, and near-end single-talk scenarios, respectively. The metrics ECHO and DEG represent the echo ratings and speech degradation ratings. }
\small
\resizebox{\columnwidth}{!}{
\begin{tabular}{c|c|c|c|c|c|c|c}
    \hline
    \multicolumn{2}{c|}{\multirow{2}{*}{Methods}}& \multicolumn{2}{c|}{FST}& \multicolumn{2}{c|}{DT}& NST& \multirow{2}{*}{Para.(M)} \\ 
    \cline{3-7}
    \multicolumn{2}{c|}{} & ERLE& ECHO& ECHO& DEG& DEG&  \\
    \hline
    \multicolumn{2}{c|}{Unprocessed}& -& 2.277& 2.607& \textbf{3.637}& 3.891& - \\
    \hline
    \multicolumn{2}{c|}{SpeexDSP}& 5.173& 3.219& 3.143& 3.443& 3.906& - \\
    \hline
    \multicolumn{2}{c|}{NSNet}& 18.964& 3.797& 3.691& 2.799& 3.873& 1.30 \\
    \hline
    \multirow{3}{*}{\makecell[c]{DTLN \\ -AEC}}& S& 29.459& 4.111& 3.508& 3.239& 3.812& 1.8 \\
    & M& 29.722& 4.152& 3.623& 3.295& 3.876& 3.9 \\
    & L& 31.990& 4.205& 3.860& 3.409& 3.878& 10.4 \\
    \hline
    \multicolumn{2}{c|}{DCCRN-AEC}& 23.871& 3.758& 4.002& 3.359& 3.943& 3.7 \\
    \hline
    \multirow{3}{*}{Ours}& S& 35.248& 4.213& 4.188& 3.242& 3.936& 1.30 \\
    & M& \textbf{38.731}& \textbf{4.311}& 4.235& 3.327& 3.935& 2.50 \\
    & L& 33.729& 4.228& \textbf{4.267}& 3.463& \textbf{3.976}& 3.61 \\
    \hline
\end{tabular}
}
\label{table_comparison}
\end{table}

We use real-recorded test set to verify the robustness of the proposed model and compare with other methods, including the conventional algorithm SpeexDSP\footnote{https://github.com/xiongyihui/speexdsp-python} and other DNN-based end-to-end methods. NSNet and DTLN-AEC \cite{westhausen2021acoustic} are the baseline model and one of the top-5 models at AEC challenge ICASSP 2021, respectively. We use the official released model of NSNet\footnote{https://github.com/microsoft/AEC-Challenge} and DTLN-AEC\footnote{https://github.com/breizhn/DTLN-aec} to do inference directly. We also modify DCCRN \cite{hu2020dccrn} to support the microphone and far-end inputs for the AEC task. The released code\footnote{https://github.com/huyanxin/DeepComplexCRN} is used to train DCCRN-AEC for 200 epochs with a batch size of 200. All the other training parameters are the same as in \cite{hu2020dccrn}. The model with the minimum validation loss is selected for testing. The results are in Table \ref{table_comparison}, from which we know that: (1) conventional AEC method like SpeexDSP tends to have series echo residues in the near-end speech. This is mostly because of its limited nonlinear modeling capacity; (2) DNN-based methods like DTLN-AEC can better suppress the echo but also degrade the quality of near-end speech, especially for the double-talk scenario. This indicates that these models may not be sensitive enough in modeling targeted speech related characteristics; (3) the model with DKG module has better balance between the echo cancellation and near-end speech retention, in both single-talk and double-talk scenarios, comparing with other methods above.

\section{Conclusions}

In this letter, we propose a DKG module that can be introduced as a learnable plug-in to the DNN model for adaptively capturing time-variant components for real-time speech enhancement. For each input audio frame, the DKG module generates an adaptive kernel to recalibrate the latent features to get the corresponding enhanced output frame. This adaptive mechanism enables the model to dynamically adjust its weights according to the input signal during inference. 
Experimental results show that introducing DKG module helps the model to dynamically and adaptively capture the speech-related characteristics in a time-variant system. 

\balance
\bibliographystyle{ieeetr}
\bibliography{AEC_references}

\end{document}